\documentclass[floatfix,reprint,amsmath,amssymb,aps,pra]{revtex4-1}
\usepackage{times,mathptmx}
\usepackage{graphicx}
\usepackage[english]{babel}
\usepackage{epstopdf}
\usepackage{color}

\begin{document}

\title{CPA-laser effect and exceptional points in $\mathcal{PT}$-symmetric multilayer structures}
\author{Denis~V.~Novitsky $^{1,2}$}
\email{dvnovitsky@gmail.com}

\affiliation{$^1$B.~I.~Stepanov Institute of Physics, National
Academy of Sciences of Belarus, Nezavisimosti Avenue 68, 220072
Minsk, Belarus \\ $^2$ITMO University, Kronversky Prospekt 49,
197101 St. Petersburg, Russia}

\begin{abstract}
The simultaneous existence of coherent perfect absorption (CPA) and lasing is one of the most intriguing features of non-Hermitian photonics. However, the link between CPA lasing and $\mathcal{PT}$ symmetry breaking at the exceptional point (EP) need clarification. In this paper, we study the manifestations of CPA-laser effect in $\mathcal{PT}$-symmetric multilayer loss-gain structure using both transfer-matrix method and numerical simulations of the Maxwell-Bloch equations. We show that the maximal contrast between absorption and amplification at different phase relations between the input waves is reached well below the EP and, therefore, is not connected to true lasing. In this regime, there is a good qualitative agreement between both computational approaches. Above the EP, the system demonstrates lasing regardless the parameters of the input waves. Thus, the maximal contrast between the absorption and amplification corresponds rather to the CPA amplifier than to the CPA laser.
\end{abstract}

\maketitle

\section{Introduction}

Non-Hermitian optics is a remarkable concept allowing to look at loss and gain in optical systems from a different, somewhat unexpected point of view. This concept has turned out to be extremely fruitful in active photonics and generated the multitude of effects in many systems. One of the most popular implementations of optical non-Hermiticity is to use the so-called $\mathcal{PT}$-symmetric structures characterized by the permittivity distribution invariant with respect to both parity and time inversion \cite{Zyablovsky2014,Feng2017,El-Ganainy2018}. Not pretending to name all interesting properties of such systems, we mention the observation in $\mathcal{PT}$-symmetric structures of the asymmetric light transmission and beam power oscillations \cite{Makris2008}, anisotropic transmission resonances \cite{Ge2012}, unidirectional ``invisibility'' \cite{Lin2011},
negative refraction and focusing \cite{Fleury2014}, topologically protected bound states \cite{Weimann2017}, light stopping \cite{Goldzak2018}, Talbot effect \cite{Ramezani2012,Wang2018}, etc.

Among many publications on optical $\mathcal{PT}$ symmetry, perhaps, the most intriguing are those devoted to simultaneous existence of lasing and antilasing in such systems. The idea of antilaser, or coherent perfect absorber (CPA), was proposed in 2010 by Chong \textit{et al.} \cite{Chong2010}. The CPA considered as a time-reversed version of a laser is based on using both absorption (instead of gain) and interference to trap the incident radiation. The idea was soon generalized by Longhi \cite{Longhi2010} who shown that both CPA and laser can be realized in the same $\mathcal{PT}$-symmetric multilayer containing balanced loss and gain slabs. In fact, the interference plays important part in this case as well allowing either to fully absorb two incoming waves or to generate two outgoing waves. As demonstrated experimentally \cite{Schindler2012,Wong2016}, the key parameter of this scheme is the phase difference $\Delta \phi$ between the waves incident on the $\mathcal{PT}$-symmetric multilayer from both sides: e.g., for $\Delta \phi=\pi/2$, one has coherent amplification of the waves, whereas they are coherently absorbed, if $\Delta \phi=-\pi/2$ (the dependence on phase difference can be reversed for the inverse order of layers in the structure). Later developments in this field allowed to demonstrate the CPA-laser effect in other types of $\mathcal{PT}$-symmetric structures, such as microrings \cite{Longhi2014}, plasmonic cavity \cite{Baum2015}, coupled resonators \cite{Sun2014}, and graphene-containing multilayers \cite{Sarisaman2018}. The existence of laser-absorber modes was also connected to the broken-symmetry phase which exists above the exceptional point \cite{Chong2011}.

We should note that the strict $\mathcal{PT}$ symmetry is not a necessary condition for lasing/antilasing effect. Other loss-gain profiles are also possible, if they provide proper distribution of loss-gain and interference of electromagnetic field resulting in enhanced absorption or amplification. The examples of such non-$\mathcal{PT}$-symmetric CPA-lasers include the so-called zero-index media \cite{Bai2016}, purely imaginary metamaterials \cite{Fu2017}, and systems with the generalized PT symmetry \cite{Sakhdari2018}.

Although there are many studies of CPA-laser effect, it is still not clear how it is connected to the exceptional point (EP) of the structure, i.e., the parameter set at which $\mathcal{PT}$-symmetry breaking occurs. It is worth noting that there are different definitions of the EP according to different notations of the scattering matrix. One of the definitions implies that the EP coincides with the point of unitary transmission \cite{Lin2011}. In this case, the CPA-laser effect is perhaps not connected to the EP position  \cite{Schindler2012}. However, this definition is problematic as shown by Ge \textit{et al.} \cite{Ge2012} who advanced another one and demonstrated its advantages in explanation of the symmetry-breaking conditions. Using this definition, it was shown that the CPA laser can be observed above the EP \cite{Chong2011}. On the other hand, the experimental verification of the CPA lasing was performed just below the EP \cite{Wong2016}. Further, we adopt Ge \textit{et al.}'s definition of the EP. It is also indirectly supported by our recent calculations in the framework of resonant loss and gain \cite{Novitsky2018}.

In this paper, we apply the resonant description of both loss and gain \cite{Novitsky2018} to the analysis of coherent absorption and amplification in $\mathcal{PT}$-symmetric multilayers. This approach based on numerical simulations of the Maxwell-Bloch equations allows to self-consistently describe \textit{dynamics} of both light field and two-level loss-gain media. Using both this approach and the stationary transfer matrix method, we analyze the conditions for CPA-laser effect in the multilayer illuminated from both sides by the counter-propagating plane monochromatic waves. For this geometry, which corresponds to the experimentally studied one \cite{Wong2016}, we calculate the output coefficient and the contrast ratio between the absorption and amplification and study their dependence on the phase difference and amplitude ratio of the waves. We assume that the CPA-laser effect corresponds to the maximal contrast between absorption and amplification under changing phase difference and leaving the other parameters (such as loss-gain level) unaltered.

Our aim is to find out whether the optimal contrast between the absorption and amplification can be associated with the EP (in our case, it is the value of pump, or imaginary part of the permittivity, where the $\mathcal{PT}$ symmetry gets broken). We show that the maximal contrast can be reached well below the exceptional point and, hence, does not require the breaking of $\mathcal{PT}$ symmetry. Above the phase transition point, lasing occurs for any phase of the input waves. Our results mean that one cannot directly associate the conditions of maximal contrast between absorption and amplification regimes with the EP and, hence, lasing \textit{per se}. It's true that there is a possibility to reach both CPA effect and lasing in the same $\mathcal{PT}$-symmetric structure, but these effects can be reached at \textit{different levels of pump}. If we take \textit{the same pumping} and change only the phase difference between the incoming waves, then the maximal contrast can be reached well below the EP. Thus, the maximal contrast corresponds rather to switching between absorption and amplification, not lasing, which can be reached only above the EP.

\section{Theoretical description of resonant loss and gain}

As proposed in our recent paper \cite{Novitsky2018}, we describe both loss and gain as a homogeneously-broadened two-level medium. Then, the Maxwell-Bloch equations for the microscopic polarization amplitude $\rho$, population difference of ground and excited states $w$ and electric field amplitude $A$ can be written as \cite{Novitsky2011}
\begin{eqnarray}
\frac{d\rho}{d\tau}&=& i l \Omega w + i \rho \delta - \gamma_2 \rho, \label{dPdtau} \\
\frac{dw}{d\tau}&=&2 i (l^* \Omega^* \rho - \rho^* l \Omega) -
\gamma_1 (w-w_{eq}),
\label{dNdtau} \\
\frac{\partial^2 \Omega}{\partial \xi^2}&-& n_d^2 \frac{\partial^2
\Omega}{\partial \tau^2}+2 i \frac{\partial \Omega}{\partial \xi}+2
i n_d^2 \frac{\partial \Omega}{\partial
\tau} + (n_d^2-1) \Omega \nonumber \\
&&=3 \alpha l \left(\frac{\partial^2 \rho}{\partial \tau^2}-2 i
\frac{\partial \rho}{\partial \tau}-\rho\right), \label{Maxdl}
\end{eqnarray}
where $\tau=\omega t$ and $\xi=kz$ are respectively the dimensionless time and distance, $\Omega=(\mu/\hbar \omega) A$ is the normalized Rabi frequency, $k = \omega/c$ is the wavenumber in vacuum, $c$ is the speed of light, $\hbar$ is the reduced Planck constant, $\mu$ is the dipole moment of the quantum transition, $\delta=(\omega_0-\omega)/\omega$ is the detuning of the light
frequency $\omega$ from the resonance frequency $\omega_0$. The dimensionless parameter $\alpha= \omega_L / \omega = 4 \pi \mu^2 C/3 \hbar \omega$ is the strength of light-matter coupling, where $\omega_L$ is the Lorentz frequency and $C$ is the concentration of two-level particles. The normalized relaxation rates of population $\gamma_1=1/(\omega T_1)$
and polarization $\gamma_2=1/(\omega T_2)$ are expressed by means of
the longitudinal $T_1$ and transverse $T_2$ relaxation times.
The local-field enhancement factor $l=(n_d^2+2)/3$ takes into account the influence of the polarization of the host dielectric with real-valued refractive index $n_d$ on the embedded active particles \cite{Crenshaw2008,Bloembergen}.

In the stationary approximation, one can obtain the effective
permittivity of a two-level medium. At the exact resonance
$\delta=0$ (this condition holds througout the paper) and for low-intensity external radiation,
$|\Omega| \ll\Omega_{sat}=\sqrt{\gamma_1 (\gamma_2^2+\delta^2)/4l^2
\gamma_2}$, the final expression is \cite{Novitsky2017}
\begin{eqnarray}
\varepsilon_{eff} \approx n_d^2+3 i l^2 \omega_L T_2 w_{eq}.
\end{eqnarray}
From this equation, one can easily see that the equilibrium population difference $w_{eq}$ is the key parameter, which allows to describe
both gain and loss materials with the Maxwell-Bloch equations
(\ref{dPdtau})-(\ref{Maxdl}). The value and sign of this parameter is governed by the external pump and, therefore, it can be called a pumping parameter. Indeed, when it is positive, we have the case of absorbing medium corresponding to low pumping. On the contrary, if $w_{eq}$ is negative, this is the case of gain medium with strong external pumping. The negativity of the equilibrium population difference in Eq. (\ref{dNdtau}) means that the external excitation tends to invert the medium and place more particles to the excited level than are on the ground one. As is well-known, the pumping cannot be fully described in the framework of the two-level model and requires consideration of other levels of the quantum particles. However, since we do not deal with the pumping processes (such as pump depletion), the two-level approximation with phenomenological account of pumping is enough for calculation of light propagation through the medium with gain already created on the transition between the two levels of interest. The two-level approach to amplifying media is well-known in laser physics \cite{Svelto}, including the use of the $w_{eq}$-like values to take pump into account \cite{Dorofeenko2012,Harayama2005}.

As shown in Ref. \cite{Novitsky2018}, it is straightforward to compose a $\mathcal{PT}$-symmetric structure
from alternating layers with balanced loss
($\varepsilon_{eff+}$) and gain ($\varepsilon_{eff-}$), where
\begin{eqnarray}
\varepsilon_{eff\pm} \approx n_d^2 \pm 3 i l^2 \omega_L T_2
|w_{eq}|. \label{epsPT}
\end{eqnarray}
Since the magnitude of the pumping parameter is the same for loss and gain layers, the necessary condition for $\mathcal{PT}$ symmetry
$\varepsilon(z) = \varepsilon^\ast(-z)$ is fulfilled, providing even
(odd) function of $z$ for the real (imaginary) part of the permittivity.

Further, we first employ the transfer-matrix method (TMM) with Eq. (\ref{epsPT}) for the permittivities of loss and gain layers to obtain the main conditions for a CPA-laser. Then we compare the TMM results with the numerical simulations of the full set of Eqs. (\ref{dPdtau})--(\ref{Maxdl}) which are solved with the finite-difference approach developed in our previous publication \cite{Novitsky2009}. As an initial value of the population difference, we employ its equilibrium value, i.e., $w(t=0)=w_{eq}$.

In this paper, we use semiconductor doped with quantum dots as an active material with the following parameters \cite{Palik,Diels}: $n_d=3.4$, $\omega_L=10^{11}$ s$^{-1}$, $T_1=1$ ns, and $T_2=0.5$ ps. The estimate of the gain coefficient $g=4 \pi \textrm{Im}(\sqrt{\varepsilon_{eff-}})/\lambda \lesssim 10^4$ cm$^{-1}$ for $\lambda \sim 1.5$ $\mu$m and $|w_{eq}| \lesssim 0.2$ shows that it can be realized in practice \cite{Babicheva12}. This choice of materials is not unique, since the multilayer parameters and light wavelength can be easily adjusted to obtain similar results with different materials. The multilayer structure contains $N=20$ unit cells with both loss and gain layers having the same thickness $d=1$ $\mu$m. 

\section{CPA laser via transfer-matrix calculations}

\begin{figure}[t!]
\includegraphics[scale=1., clip=]{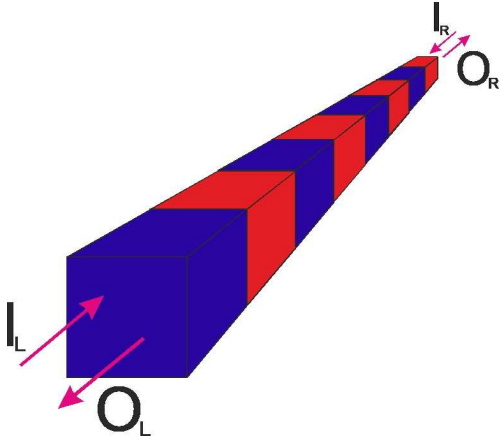}
\caption{\label{fig1} Schematic of the multilayered $\mathcal{PT}$-symmetric structure under consideration. Blue color indicates loss layers, whereas red color is for gain ones.}
\end{figure}

\begin{figure}[t!]
\includegraphics[scale=1., clip=]{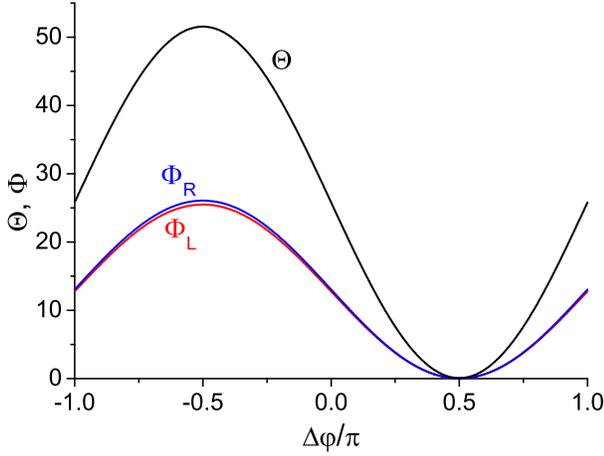}
\caption{\label{fig2} Dependence of the output coefficient $\Theta$ on the phase difference $\Delta \varphi$. Other parameters: $|w_{eq}|=0.2$, $\sigma=1$.}
\end{figure}

\begin{figure}[t!]
\includegraphics[scale=1., clip=]{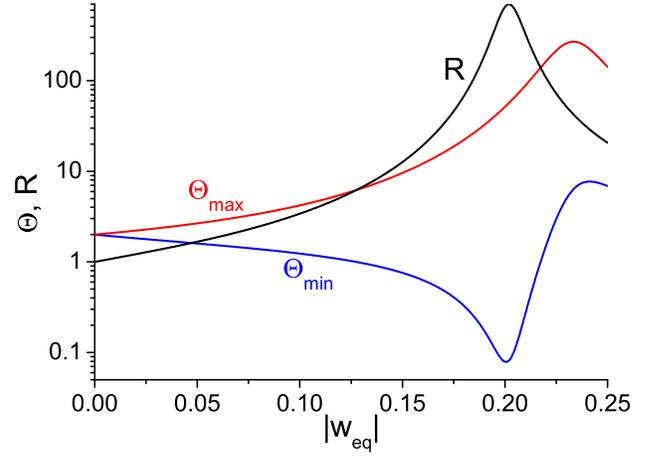}
\caption{\label{fig3} Dependence of the maximal and minimal output coefficient $\Theta$ (at $\Delta \varphi=-\pi/2$ and $\pi/2$, respectively) and the contrast ratio $R$ on the pumping parameter $|w_{eq}|$. Other parameters: $\sigma=1$.}
\end{figure}

\begin{figure}[t!]
\includegraphics[scale=1., clip=]{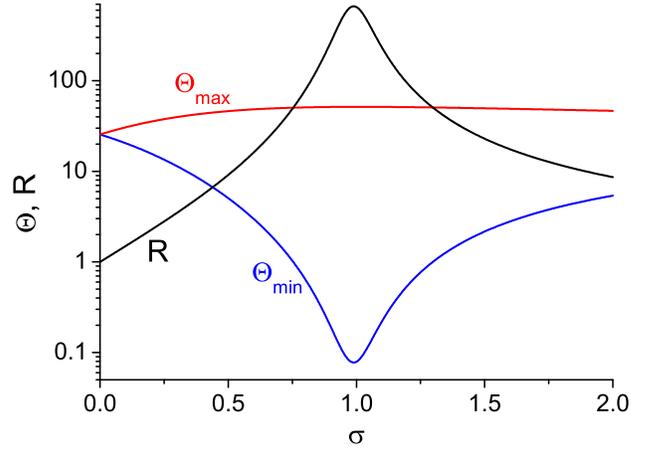}
\caption{\label{fig4} Dependence of the maximal and minimal output coefficient $\Theta$ (at $\Delta \varphi=-\pi/2$ and $\pi/2$, respectively) and the contrast ratio on the ratio of incident eaves amplitudes $\sigma$. Other parameters: $|w_{eq}|=0.2$.}
\end{figure}

The scheme of the one-dimensional loss-gain multilayer is shown in Fig. \ref{fig1}. To excite the gain layers, one can employ the side pumping scheme similar to that realized in Ref. \cite{Wong2016}. It was shown previously \cite{Novitsky2018} that the structure with the parameters given above demonstrates the characteristic features of $\mathcal{PT}$-symmetric system, such as anisotropic transmission resonances and symmetry-breaking phase transition (at $|w_{eq}|>0.22$). Those results were obtained for the single input monochromatic wave with $\lambda=1.513$ $\mu$m. For the CPA-laser effect, it is of fundamental importance to have two coherent input waves from both sides of the structure, since the key ingredient is the interference between the phase-shifted waves. In this section, we analyze the main conditions for CPA-laser by using TMM calculations with the stationary permittivities Eq. (\ref{epsPT}). 

Both incident waves are assumed to have the same wavelength $\lambda=1.513$ $\mu$m and are shifted in phase by $\Delta \varphi$, so that the ratio of field strengths for right- and left-incident fields is given by $E_{right}/E_{left}=\sigma e^{i \Delta \varphi}$, where $\sigma$ is the real number showing the ratio of field amplitudes. This means that the total input intensity is $I=I_L+I_R=I_L(1+\sigma^2)$. The left-output field is formed as a sum of reflection of the left-incident wave $r_L$ and transmission of the right-incident one $t_R$. Analogous condition is valid for the right-output field. The amplitude reflection and transmission coefficients can be easily expressed through corresponding elements of the structure's transfer matrix, so that we can write the formulas for the output intensities as follows:
\begin{eqnarray}
O_L=|r_L+t_R|^2=I_L \left| \frac{M_{21}+||M|| \sigma e^{i \Delta \varphi}}{M_{11}} \right|^2 = I_L \Phi_L, \label{intleft} \\
O_R=|t_L+r_R|^2=I_L \left| \frac{1-M_{12} \sigma e^{i \Delta \varphi}}{M_{11}} \right|^2= I_L \Phi_R. \label{intright}
\end{eqnarray}
The transfer matrix of a multilayer structure $M$ can be obtained in a standard way, see, e.g., Ref. \cite{BornWolf}. Since we deal with the case of normal incidence, the transfer matrix of the multilayer can be represented in especially simple form \cite{Novitsky2008},
\begin{eqnarray}
M=\Delta_{01}(\Pi_1 \Delta_{12} \Pi_2 \Delta_{21})^N \Delta_{10}, \nonumber
\end{eqnarray}
where the matrices $\Delta_{lm}$ and $\Pi_l$ describe the reflection and propagation of light, respectively:
\begin{eqnarray}
\Delta_{lm}&=&\left(\begin{array}{cc} {\delta_{lm}^+} &
{\delta_{lm}^-}
\\ {\delta_{lm}^-} & {\delta_{lm}^+} \end{array} \right), \qquad
\delta_{lm}^\pm = \frac{1}{2} (1 \pm \frac{n_m}{n_l}),
\nonumber \\
\Pi_l&=&\left(\begin{array}{cc} {\exp^{-i n_l d \omega/c}} & {0}
\\ {0} & {\exp^{i n_l d \omega/c}} \end{array} \right). \nonumber
\end{eqnarray}
Here $n_i$ is the refractive index of the $l$th layer ($l=1,2$), $n_0$ is the refractive index of the ambient medium. $||M||$ is the determinant of the transfer matrix. The total output intensity is $O=O_L+O_R$.

As a main parameter, we use the output coefficient of the CPA-laser \cite{Wong2016}: 
\begin{eqnarray}
\Theta=2 \frac{O}{I} = 2 \frac{\Phi_L+\Phi_R}{1+\sigma^2}. \label{output}
\end{eqnarray}
A factor $2$ means that Eq. \ref{output} gives the output
intensity per one input channel. We search for the conditions, when $\Theta$ reaches minimum (CPA) and maximum (lasing). The contrast ratio between these maxima and minima
\begin{eqnarray}
R=\Theta_{max}/\Theta_{min} \label{contrast}
\end{eqnarray}
shows the intensity contrast between the two regimes reached in the CPA-laser.

Figure \ref{fig2} shows the dependence of the output intensities $\Phi_{L,R}$ and the output coefficient $\Theta$ on the phase difference $\Delta \varphi$ between the waves of the same amplitude ($\sigma=1$); the pumping parameter is assumed to be $|w_{eq}|=0.2$. It is readily seen that the minimal value of the output coefficient $\Theta_{min}<<1$ is reached exactly at $\Delta \varphi=\pi/2$, whereas the maximum $\Theta_{max}>50$ is at $\Delta \varphi=-\pi/2$. This is in full accordance with Ref. \cite{Wong2016}. This effect can be understood as a consequence of interference, so that changing the phase difference $\Delta \varphi$ from $\pi/2$ to $-\pi/2$ results in switching maxima of light intensity from loss layers (CPA) to gain layers (amplification).

Let us study how the quantities $\Theta_{min}$ (calculated at fixed $\Delta \varphi=\pi/2$), $\Theta_{max}$ (calculated at fixed $\Delta \varphi=-\pi/2$) and the contrast ratio $R$ vary with the pumping parameter $|w_{eq}|$ and the amplitude ratio $\sigma$. The dependencies on $|w_{eq}|$ (at $\sigma=1$) are demonstrated in Fig. \ref{fig3}. $\Theta_{max}$ increases with growing pump and reaches the maximum at $|w_{eq}| \approx 0.23$ (just above the EP), whereas $\Theta_{min}$ decreases, has the minimal value at $|w_{eq}| \approx 0.20$ and then rapidly grows approaching the EP. As a result, the peak value of $R$ (up to about $700$) occurs at the same $|w_{eq}| \approx 0.20$, where the dip of $\Theta_{min}$ occurs. This means that the optimal (from the contrast maximization point of view) value of pumping is reached significantly below the EP and cannot be attributed simply to the effects of $\mathcal{PT}$ symmetry breaking (such as onset of lasing \cite{Novitsky2018}).

\begin{figure}[t!]
\includegraphics[scale=1., clip=]{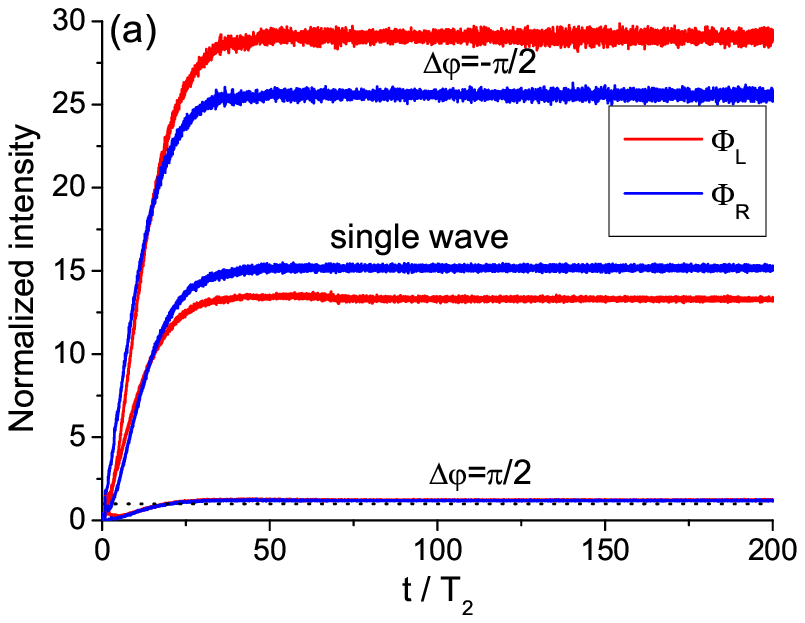}
\includegraphics[scale=0.95, clip=]{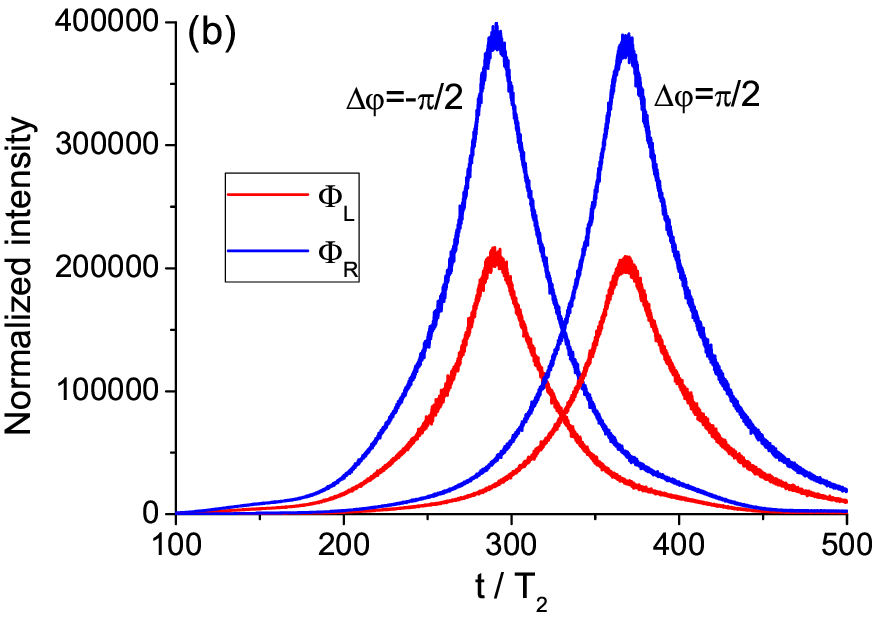}
\caption{\label{fig5} Temporal dynamics of the output intensities at the phase difference $\Delta \varphi=-\pi/2$ and $\pi/2$ at the pumping parameters (a) $|w_{eq}|=0.20$ and (b) $|w_{eq}|=0.24$. Other parameters: $\sigma=1$. For comparison, in panel (a), the case of single wave incident from left or right is shown.}
\end{figure}

\begin{figure}[t!]
\includegraphics[scale=1., clip=]{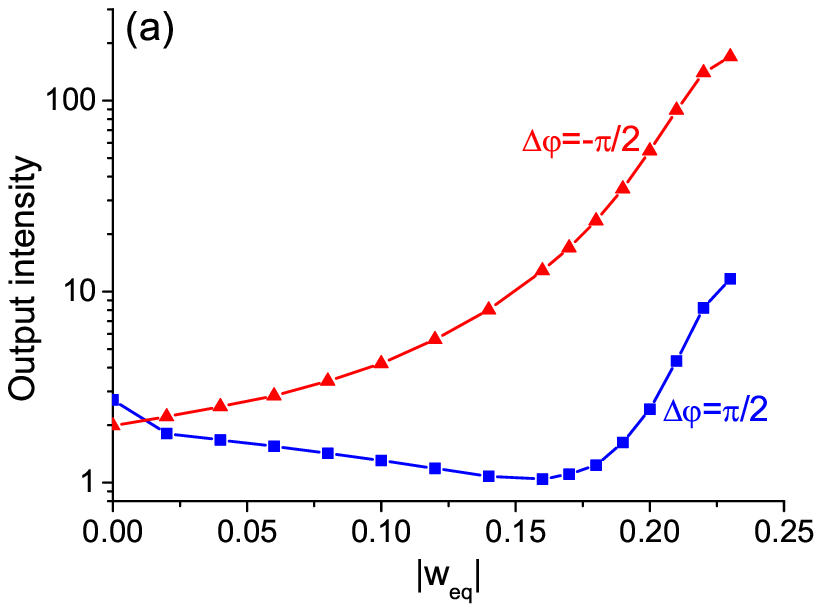}
\includegraphics[scale=1., clip=]{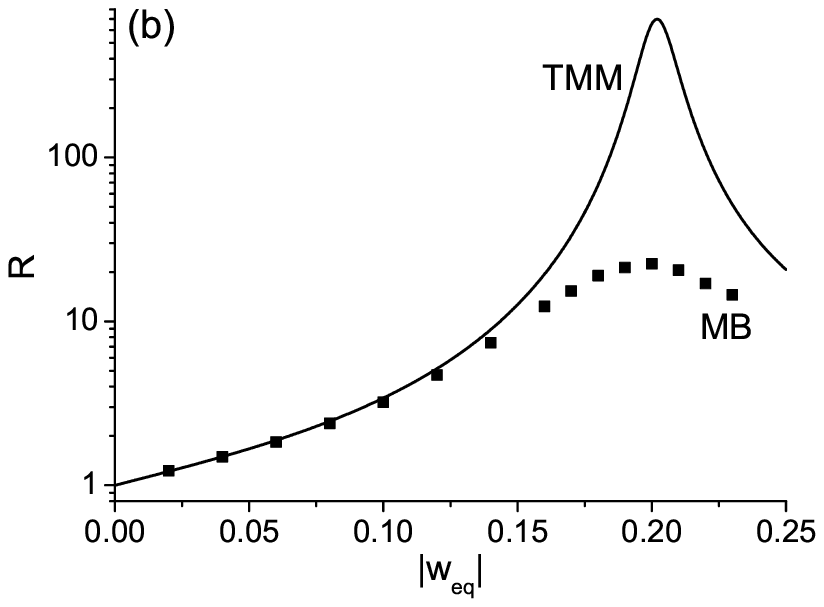}
\caption{\label{fig6} Dependence of (a) the output coefficient $\Theta$ (at $\Delta \varphi=-\pi/2$ and $\pi/2$, respectively) and (b) the contrast ratio $R$ on the pumping parameter $|w_{eq}|$. Other parameters: $\sigma=1$. For comparison, the values of $R$ are given from the TMM and Maxwell-Bloch (MB) calculations.}
\end{figure}

Similar analysis can be performed for the dependence on the amplitude ratio $\sigma$ shown in Fig. \ref{fig4} (at $|w_{eq}|=0.2$). $\Theta_{max}$ has very weak dependence on $\sigma$, therefore the contrast $R$ is fully determined by the behavior of $\Theta_{min}$. As it could be expected, the maximum of $R$ corresponds to the symmetric situation of two waves with equal amplitudes ($\sigma=1$). It is also worth mentioning that the asymmetry ($\sigma \neq 1$) does not change the phase dependence shown in Fig. \ref{fig2}, but shifts the position of $R$ maximum and the $\Theta_{min}$ dip along the $|w_{eq}|$ axis. This can be viewed as an instrument for tuning the $R$ peak position with respect to the EP -- closer ($\sigma<1$) or farther ($\sigma>1$) from it.

Note that the curves for $R$ in Figs. \ref{fig3} and \ref{fig4} have very narrow resonance (this may be not obvious due to logarithmic scale). The same is true, if we plot the spectral dependence changing $\lambda$ and leaving all the parameters of the media unaltered (not shown here). Though this procedure (which gives the peak at our chosen $\lambda=1.513$ $\mu$m) cannot be strictly justified (one should take into account the linewidth of the particles resonance as well), it allows to feel the importance of subtle match between the structure geometry and the wavelength to observe the optimal CPA-laser effect.

\section{CPA laser via simulations of the Maxwell-Bloch equations}

\begin{figure}[t!]
\includegraphics[scale=1., clip=]{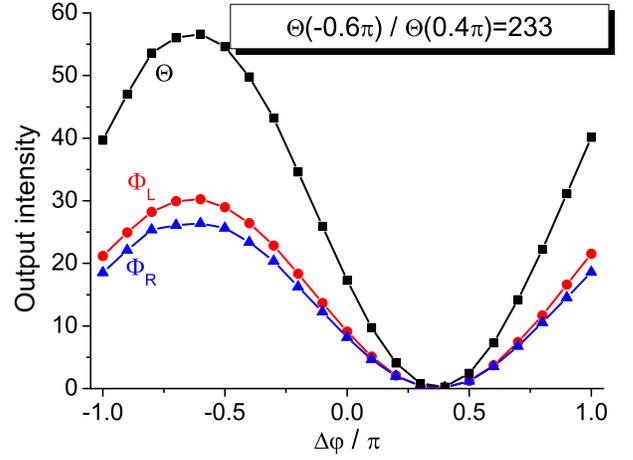}
\caption{\label{fig7} The same as in Fig. \ref{fig2}, but calculated via Maxwell-Bloch equations.}
\end{figure}

In this section, we compare the stationary analysis given above with the full numerical simulations of the Maxwell-Bloch equations. As previously, we take two counter-propagating waves of the same intensity ($\sigma=1$, the absolute amplitude is $\Omega_0=10^{-5} \gamma_2$) and the phase difference $\Delta \varphi$. The dynamics of output intensities calculated for $|w_{eq}|=0.2$ is shown in Fig. \ref{fig5}(a). One can see the rapid establishment of the stationary level of the output radiation. These dynamics are characteristic for the $\mathcal{PT}$-symmetric state. With respect to the single-wave case, the stationary output intensities are much greater for $\Delta \varphi=-\pi/2$ and much smaller for $\Delta \varphi=\pi/2$. This is in qualitative conformity with the discussion in the previous section. However, the results do no coincide quantitatively. One can see this in Fig. \ref{fig6}(a). Although the output intensity for both $\Delta \varphi=-\pi/2$ and $\Delta \varphi=\pi/2$ behaves similar to the curves in Fig. \ref{fig3}, the maximum in the first case is not so high and the minimum in the second one is not so deep. In addition, the minimum is reached at $|w_{eq}|=0.16$, not at $0.20$ as in Fig. \ref{fig3}.

We stop in Fig. \ref{fig6}(a) at the pumping parameter $|w_{eq}|=0.23$, since above this value the system jumps into the broken-symmetry state. In this latter state, the system generates powerful light pulses as shown in Fig. \ref{fig5}(b) for $|w_{eq}|=0.24$. One can see that the phase difference between the incident waves does not influence the intensity of this lasing pulses. Only the time of pulse appearance can be controlled with $\Delta \varphi$. This implies that the CPA-laser effect should be searched for only below the EP.

The contrast ratios shown in Fig. \ref{fig6}(b) corroborate that the Maxwell-Bloch simulations strongly underestimate the value of $R$ (only about $22$) in comparison to the calculations within TMM ($R \approx 700$). The possible reason is the narrowness of the resonance pointed out in the end of the previous section, so that the discretized version of the structure used in numerical simulations do not correspond perfectly to the TMM model. On the other hand, the Maxwell-Bloch simulations taking saturation into account are more reliable in the vicinity of the EP as noticed in Ref. \cite{Novitsky2018}. In addition, this method allows to calculate the temporal dynamics, which is beyond the scope of TMM. Nevertheless, the position of the contrast-ratio peak ($|w_{eq}|=0.20$) is identical according to both approaches which can be considered as complementary.

Finally, in Fig. \ref{fig7}, we plot the phase dependence of the output intensity obtained via the Maxwell-Bloch simulations at $|w_{eq}|=0.20$. Perhaps, due to the reasons discussed above, the dependence is shifted in comparison to the analogous relationship obtained within TMM (Fig. \ref{fig2}): the minimum is here observed at $\Delta \varphi=0.4 \pi$, whereas the maximum is at $\Delta \varphi=-0.6 \pi$. Taking this into account and calculating the contrast ratio for these shifted phase differences, we have $R=\Theta_{max}/\Theta_{min}=\Theta(-0.6 \pi)/\Theta(0.4 \pi) \approx 230$, which is much greater than only $22$ reported in Fig. \ref{fig6}(b) and better corresonds to the TMM values.

\section{Conclusion}

In this paper, we have analyzed the conditions for CPA-laser effect in the $\mathcal{PT}$-symmetric multilayer structure with resonant loss and gain illuminated by two counter-propagating waves. We employed two methods -- the standard transfer-matrix method in the steady-state approximation and the numerical simulations of the full set of the Maxwell-Bloch equations. The results (the pump- and phase-dependencies of the output coefficient and the contrast ratios of the maximal and minimal outputs) given by both methods are in good correspondence, in particular the position of the contrast-ratio peak is reliably determined. The quantitative discrepancy between the approaches is perhaps due to the narrow spectral resonance and the proximity to the EP. We should emphasize that according to our calculations, the maximum of the contrast ratio in the case of equal-amplitude incident waves is located well below the EP and, hence, does not requires $\mathcal{PT}$ symmetry breaking and lasing \textit{per se}. Therefore, it would be more correct to say about CPA-amplifier, but not CPA-laser in these conditions. We believe that our results will be helpful to clarify the properties of $\mathcal{PT}$-symmetric or similar loss-gain structures.

\acknowledgements{The author is grateful to Viktoryia Kouhar for help in figures preparation. The work was supported by the Belarusian State Program of Scientific Research ``Photonics, Opto- and Microelectronics'' (Task 1.2.02) and the Russian Foundation for Basic Research (Projects No. 18-32-00160 and 18-02-00414). Numerical simulations of Maxwell-Bloch equations were supported by the Russian Science Foundation (Project No. 18-72-10127).}


\begin{thebibliography}{00}

\bibitem{Zyablovsky2014} A.~A.~Zyablovsky, A.~P.~Vinogradov, A.~A.~Pukhov, A.~V.~Dorofeenko, and A.~A.~Lisyansky, 
\textit{Phys. Usp.} \textbf{57}, 1063 (2014).

\bibitem{Feng2017} L.~Feng, R.~El-Ganainy, and L.~Ge,
{Nat. Photon.} \textbf{11}, 752 (2017).

\bibitem{El-Ganainy2018} R.~El-Ganainy, K.~G.~Makris, M.~Khajavikhan, Z.~H.~Musslimani, S.~Rotter, and D.~N.~Christodoulides,
{Nat. Phys.} \textbf{13}, 11 (2018).

\bibitem{Makris2008} K.~G.~Makris, R.~El-Ganainy, D.~N.~Christodoulides, and Z.~H.~Musslimani, 
\textit{Phys. Rev. Lett.} \textbf{100}, 103904 (2008).

\bibitem{Ge2012} L.~Ge, Y.~D.~Chong, and A.~D.~Stone, 
\textit{Phys. Rev. A} \textbf{85}, 023802 (2012).

\bibitem{Lin2011} Z.~Lin, H.~Ramezani, T.~Eichelkraut, T.~Kottos, H.~Cao, and D.~N.~Christodoulides, 
\textit{Phys. Rev. Lett.} \textbf{106}, 213901 (2011).

\bibitem{Fleury2014} R.~Fleury, D.~L.~Sounas, and A.~Al\`{u},
\textit{Phys. Rev. Lett.} \textbf{113}, 023903 (2014).

\bibitem{Weimann2017} S.~Weimann, M.~Kremer, Y.~Plotnik, Y.~Lumer,
S.~Nolte, K.~G.~Makris, M.~Segev, M.~C.~Rechtsman, and A.~Szameit,
\textit{Nat. Mater.} \textbf{16}, 433 (2017).

\bibitem{Goldzak2018} T.~Goldzak, A.~A.~Mailybaev, and N.~Moiseyev,
\textit{Phys. Rev. Lett.} \textbf{120}, 013901 (2018).

\bibitem{Ramezani2012} H.~Ramezani, D.~N.~Christodoulides, V.~Kovanis, I.~Vitebskiy, and T.~Kottos,
\textit{Phys. Rev. Lett.} \textbf{109}, 033902 (2012).

\bibitem{Wang2018} S.~Wang, B.~Wang, and P.~Lu,
{Phys. Rev. A} \textbf{98}, 043832 (2018).

\bibitem{Chong2010} Y.~D.~Chong, L.~Ge, H.~Cao, and A.~D.~Stone,
{Phys. Rev. Lett.} \textbf{105}, 053901 (2010).

\bibitem{Longhi2010} S.~Longhi,
{Phys. Rev. A} \textbf{82}, 031801(R) (2010).

\bibitem{Schindler2012} J.~Schindler, Z.~Lin, J.~M.~Lee, H.~Ramezani, F.~M.~Ellis, and T.~Kottos,
{J. Phys. A} \textbf{45}, 444029 (2012).

\bibitem{Wong2016} Z.~J.~Wong, Y.-L.~Xu, J.~Kim, K.~O'Brien, Y.~Wang, L.~Feng, and X.~Zhang,
{Nat. Photon.} \textbf{10}, 796 (2016).

\bibitem{Longhi2014} S.~Longhi,
{Opt. Lett.} \textbf{39}, 5026 (2014).

\bibitem{Baum2015} B.~Baum, H.~Alaeian, and J.~Dionne,
{J. Appl. Phys.} \textbf{117}, 063106 (2015).

\bibitem{Sun2014} Y.~Sun, W.~Tan, H.~Q.~Li, J.~Li, and H.~Chen,
{Phys. Rev. Lett.} \textbf{112}, 143903 (2014).

\bibitem{Sarisaman2018} M.~Sarısaman and M.~Tas,
{J. Opt. Soc. Am. B} \textbf{35}, 2423 (2018).

\bibitem{Chong2011} Y.~D.~Chong, L.~Ge, and A.~D.~Stone,
{Phys. Rev. Lett.} \textbf{106}, 093902 (2011).

\bibitem{Bai2016} P.~Bai, K.~Ding, G.~Wang, J.~Luo, Z.-Q.~Zhang, C.~T.~Chan, Y.~Wu, and Y.~Lai,
{Phys. Rev. A} \textbf{94}, 063841 (2016).

\bibitem{Fu2017} Y.~Fu, Y.~Cao, S.~A.~Cummer, Y.~Xu, and H.~Chen, 
{Phys. Rev. A} \textbf{96}, 043838 (2017).

\bibitem{Sakhdari2018} M.~Sakhdari, N.~M.~Estakhri, H.~Bagci, and P.-Y.~Chen,
{Phys. Rev. Appl.} \textbf{10}, 024030 (2018).

\bibitem{Novitsky2018} D.~V.~Novitsky, A.~Karabchevsky, A.~V.~Lavrinenko, A.~S.~Shalin, and A.~V.~Novitsky,
{Phys. Rev. B} \textbf{98}, 125102 (2018).

\bibitem{Novitsky2011} D.~V.~Novitsky, 
{Phys. Rev. A} \textbf{84}, 013817 (2011).

\bibitem{Crenshaw2008} M.~E.~Crenshaw, 
{Phys. Rev. A} \textbf{78}, 053827 (2008).

\bibitem{Bloembergen} N.~Bloembergen, \textit{Nonlinear Optics} (Benjamin, New York, 1965).

\bibitem{Novitsky2017} D.~V.~Novitsky, V.~R.~Tuz, S.~L.~Prosvirnin, A.~V.~Lavrinenko, and A.~V.~Novitsky,
{Phys. Rev. B} \textbf{96}, 235129 (2017).

\bibitem{Svelto} O.~Svelto, \textit{Principles of
Lasers}, 1st ed. (Springer, New York, 1976).

\bibitem{Dorofeenko2012} A.~V.~Dorofeenko, A.~A.~Zyablovsky, A.~A.~Pukhov, A.~A.~Lisyansky, and A.~P.~Vinogradov, {Phys. Usp.} \textbf{55}, 1080 (2012).

\bibitem{Harayama2005} T.~Harayama, S.~Sunada, and K.~S.~Ikeda, {Phys. Rev. A} \textbf{72}, 013803 (2005).

\bibitem{Novitsky2009} D.~V.~Novitsky, 
{Phys. Rev. A} \textbf{79}, 023828 (2009).

\bibitem{Palik} E.~D.~Palik (ed.), \textit{Handbook of Optical Constants of Solids} (Academic Press, San Diego, 1998).

\bibitem{Diels} J.-C.~Diels and W.~Rudolph, \textit{Ultrashort Laser Pulse Phenomena}, 2nd ed. (Academic Press, San Diego, 2006).

\bibitem{Babicheva12} V.~E.~Babicheva, I.~V.~Kulkova, R.~Malureanu, K.~Yvind, and A.~V.~Lavrinenko,
\textit{Photon. Nanostruct. -- Fund. Appl.} \textbf{10}, 389 (2012).

\bibitem{BornWolf} M.~Born and E.~Wolf, \textit{Principles of Optics}, 7th ed. (Cambridge University Press, Cambridge, 1999).

\bibitem{Novitsky2008} D.~V.~Novitsky and S.~Yu.~Mikhnevich, 
{J. Opt. Soc. Am. B} \textbf{25}, 1362 (2008).

\end{thebibliography}
\end{document}